\documentclass[final,times,twocolumn,authoryear]{elsarticle}

\usepackage{amssymb}
\usepackage{url}

\newcommand\op[1]{\mathop{\rm #1}\nolimits}

\newcommand\N{{\mathcal N}}

\newcommand{\weg}[1]{}

\newcommand\dd{\stackrel{\op{d}}{\sim}}
\newcommand\Like{\mathcal{L}}

\newtheorem{theoremc}{Theorem}
\newtheorem{rk}[theoremc]{Remark}

\usepackage{fullpage}
\usepackage{color}

\makeatletter
\def\ps@pprintTitle{%
  \let\@oddhead\@empty
  \let\@evenhead\@empty
  \let\@oddfoot\@empty
  \let\@evenfoot\@oddfoot
}
\makeatother

\begin{document}

\begin{frontmatter}

 \title{A multifractal approach towards inference in  finance}

\author{Ola L{\o}vsletten\corref{cor1}}
 \ead{ola.lovsletten@uit.no}
\author{Martin Rypdal}
 \address{Department of Mathematics and Statistics, University of Troms{\o}, N-9037 Troms{\o}, Norway.}

\begin{abstract}
We introduce tools for inference in the multifractal random walk introduced by \cite{Bacry:2001gl}. These tools include formulas for smoothing, filtering and volatility forecasting. In addition, we present methods for computing conditional densities for one- and multi-step returns. The inference techniques presented in this paper, including maximum likelihood estimation, are applied to data from the Oslo Stock Exchange, and it is observed that the volatility forecasts based on the multifractal random walk have a much richer structure than the forecasts obtained from a basic stochastic volatility model.  
\end{abstract}

\begin{keyword}
Multifractal \sep inference \sep volatility forecasting \sep Laplace approximation 
\end{keyword}
\end{frontmatter}

\section{Introduction} \label{intro}
Modeling financial time series by stochastic processes dates back to the work of \cite{Bachelier:1900vf}. Bachelier proposed to model the price of a financial asset as a Brownian motion with drift.
It was later realized, by e.g. \cite{Mitchell:1915vj}, that the standard deviation of price changes are proportional to the price levels themselves. Therefore, Bachelier's model should be modified so that it is the {\em logarithmic} asset price, $X(t) = \log P(t)$, that is modeled as a Brownian motion with drift. As a modification of this model, \citet{Mandelbrot:1963ui} proposed to replace Brownian motion with $\alpha$-stable L{\'e}vy processes with $\alpha<2$, so-called L{\'e}vy flights.  

Both Brownian motions and L\'evy flights are selfsimilar and have independent increments. However, empirical analyses of asset prices have revealed that, even though logarithmic returns are uncorrelated, they are nevertheless strongly dependent. This stylized fact is called volatility clustering, and it is not well described by Brownian motions nor L{\'e}vy flights. Other processes, such as stochastic volatility (SV) models, are specifically designed to include this feature. The simplest example is the basic SV model of \citet{Taylor:1982tx}. If we choose\footnote{Since $\mu$ is easily estimated from data, this can be assumed without any loss of generality.} $\mu=0$, this model is defined by the stochastic differential equation 
$$
dX(t) = \sigma (t) \,dB(t)\,,
$$ 
where the logarithmic volatility varies according to an Ornstein-Uhlenbeck process, i.e.  
\begin{equation}
d \log \sigma (t) = -a \log \sigma (t) dt + \nu d \tilde{B}(t)\,,
\end{equation}
where $\tilde{B}(t)$ is a Brownian motion independent of $B(t)$. 

Another class of SV models are the multifractal random processes. These models come from turbulence theory, and their origin can be traced back to works of \cite{Kolmogorov:1962vx} and \cite{Obukhov:1962wx}. 
The defining properties of a multifractal process $X(t)$ are stationary increments and structure functions that are power-laws in time, i.e.
\begin{equation}
\mathbb{E}[|X(t)|^q] \sim t^{\zeta(q)}\,.
\label{Skalering}
\end{equation}
The scaling functions $\zeta(q)$ are linear for selfsimilar processes, but usually the term ``multifractal'' refers to the cases where $\zeta(q)$ are strictly concave. For such processes, the absolute values of the increments of $X(t)$ may have algebraically decaying auto-correlation functions (ACFs), even though the increments themselves are uncorrelated. In contrast, the ACFs for the absolute values of the increments decay exponentially in the basic SV models.

That multifractals represent a suitable framework for modeling financial time series was first discovered about fifteen years ago by \cite{Ghashghaie:1996wa} and \cite{Mandelbrot:1997wz}. Shortly after this  \cite{Calvet:2001cw} showed how one can obtain a discrete-time SV model as a discretization  of a continuous-time multifractal.

The model constructed by Calvet and Fisher is called the Markov-Switching Multifractal (MSM), and it is constructed by randomizing the so-called multiplicative cascade. The result is a model that describes log-returns as
\begin{equation} \label{MSM}
x_t = \sigma\,\sqrt{M_t}\, \varepsilon_t\,.
\end{equation}
Here $\varepsilon_t \dd \mathcal{N}(0,1)$ are independent variables and the volatility is a product on the form
$$
M_t=M_{t,1} M_{t,2} \cdots M_{t,K}\,.
$$
The variables $M_{t,k}$ are updated with different frequencies for different levels $k$. To be precise, at each time step $t$, $M_{t,k}$ is given a new value (independently drawn from a distribution $M$) with probability $\gamma_k$ and left unchanged with probability $1-\gamma_k$. The approximate multifractality in the MSM model is achieved by choosing $\gamma_k=1-(1-\gamma_1)^{b^{k-1}}$ for some $\gamma_1 \in (0,1)$ and some $b>0$. By exploiting general techniques for Markov-Switching models, Calvet and Fisher have developed inference methods for the MSM model, including maximum likelihood (ML) estimation and volatility forecasting. Unfortunately, there are some limitations to the applicability of these methods. One problem is that the likelihood functions only are available when $M$ is discrete, something that leads to rather unnatural parameterizations. Also, in practice, it is only possible to compute the likelihood if the parameter $K$ does not exceed $\approx 10$ \citep{Lux:2008tf}. In effect, this introduces an unwanted exponential cutoff in the volatility dependence, at the time scale $b^K$.      

At the same time that Calvet and Fisher proposed the MSM model, \cite{Bacry:2001gl} presented a different type of multifractal process, the so-called multifractal random walk (MRW). A popular discrete-time approximation to this process is given by equation (\ref{MSM}), with 
\begin{equation}
M_t= c e^{h_t},
\label{M}
\end{equation}where $h_t$ is a stationary and centered Gaussian process with co-variances 
\begin{equation} \label{cov}
\op{Cov}(h_t,h_s) = \lambda^2 \log^+ \frac{\mathcal{T}}{(|t-s|+1) \Delta t}\,.
\end{equation} 
The constant $c$ is chosen so that $1/c = \mathbb{E}[e^{h_t}]$.
If the step-length $\Delta t$ is fixed\footnote{The variable $t$ is dimensionless and represents the number of time steps of length $\Delta t$.} it is convenient to denote $R= \mathcal{T}/\Delta t$. The model then depends on three parameters: $\theta=(\lambda, \sigma, R)$. 

For the purpose of modeling financial time series, an important property of the MRW model is the slow decay of the volatility dependence. Since the innovations $\varepsilon_t$ are independent, the auto-correlation function for the process $|x_t|$ becomes 
\begin{equation} \label{voldecay}
\mathbb{E}[|x_t x_s|] \propto e^{\frac12 \mathbb{E}[(h_t/2+h_s/2)^2]} \propto e^{\frac14 \op{Cov}(h_t,h_s)}\,,
\end{equation}
which for $1 \ll s \ll R$ gives the approximate scaling 
\begin{equation}
\mathbb{E}[|x_t x_{t+s}|] \sim s^{-\frac{1}{4} \lambda^2}\,.
\end{equation}    
The parameter $\lambda$ is called the intermittency parameter, and it also determines the nonlinearity of the scaling function. In fact, the scaling function of the (continuous-time) MRW model is 
$$
\zeta(q) = \frac{1}{2} \Big{(} 1+ \frac{\lambda^2}{2} \Big{)}\,q - \frac{\lambda^2}{8}\,q^2\,.
$$

In contrast to the MSM model, which is obtained by randomizing a discrete multiplicative cascade, the MRW model builds on a continuous cascade. In fact, the log-normal MRW model that we consider in this paper is just a special case of a more general class of processes known as infinitely divisible cascades \citep{Muzy:2002ua}. These processes have very desirable theoretical properties, e.g. exact multifractal scaling. From this point of view, the MRW model is preferable over the MSM model, and it is therefore important to develop inference techniques for the MRW model. A step in this direction was taken in \citep{Lovsletten:2011vm}, where we presented methods for ML estimation. These results were obtained by observing that the processes defined by equations (\ref{MSM}) and (\ref{cov}) are very similar to discrete-time versions of the basic SV models. In fact, if we replace the process $h_t$ with an auto-regressive model of order one (an AR(1) process),
\begin{equation} \label{AR}
h_t = \psi \,h_{t-1} + \sigma_u u_t\,, 
\end{equation}
where $u_t$ is Gaussian white noise with unit variance,
then the process defined by equations (\ref{MSM}) and (\ref{M}) is a basic SV model. Hence we can use existing techniques for basic SV models \citep{Skaug:2009vna,Martino:2012be}    
in combination with general ML methods for Gaussian processes \citep{McLeod:2007wp} to obtain likelihoods for the MRW model.   

While in \citep{Lovsletten:2011vm} we focused on parameter estimation, the focus of this paper is primarily conditional forecasts of returns and inference regarding the latent variables $h_t$. To be more precise, we are interested in estimating the conditional variables $h_s|\{x_t , t\leq T\}$. For $s<T$ this problem is known as smoothing, for $s=T$ it is called filtering, and for $s>T$ it is called forecasting. These techniques are of obvious importance for the applicability of the MRW model in finance. 

The paper is structured as follows. In section \ref{sv} we review inference techniques for basic SV models, and in section \ref{mrwsec} we generalize these results to the MRW model. In section \ref{example} we apply some of these methods to data from the Oslo Stock Exchange, and in section \ref{conclusion} we give some concluding remarks.

\section{Inference techniques in the basic SV model} \label{sv}
In general, many statistical problems in stochastic modeling, e.g. model selection, parameter estimation and assessment of uncertainty in estimates, can be solved by utilizing the likelihood of the model. 
Given data $z=(z_1,...,z_T)$, the likelihood $\Like$ of a random vector $x=(x_1,\dots,x_T)$, with probability density function $p_x(\cdot|\theta)$, is defined as the function
\begin{equation}
\Like(\theta|z)=p_x(z|\theta),
\end{equation}
i.e. one views the the probability density as a function of the parameters $\theta$, with $z$ fixed.  
\begin{rk}{\em 
To simplify notations we will drop the subscripts on the densities throughout the rest of the paper. It will be clear from the arguments which densities are considered. We also suppres the dependency of the parameter vector $\theta$ in the notation of the densities.}
\end{rk}

In the basic SV model the likelihoods are difficult to compute directly. By conditioning on the latent field $h=(h_1,\ldots,h_T)$, the probability density of $x$ takes the form
\begin{equation} \label{integral}
p(x) = \int_{\mathbb{R}^T} p(x|h)p(h)dh\,,
\end{equation}
where the joint density $p(x,h)=p(x|h)p(h)$ is a product of the Gaussian marginals
\begin{equation} \label{fac1}
p(x|h)=\prod_{i=1}^T p(x_i|h_i)
\end{equation}
and 
\begin{equation} \label{fac2}
 p(h)=p(h_1)\prod_{i=2}^T p(h_i|h_{i-1})\,.
\end{equation}
The factors in equations (\ref{fac1}) and (\ref{fac2}) are densities corresponding to the distributions
$x_t|h_t \dd \N(0,\sigma^2c\exp(h_t))$, 
$h_1 \dd \N(0,\sigma_u^2/(1-\psi^2))$ and 
$h_t|h_{t-1}  \dd \N(\psi h_{t-1} ,\sigma_u^2)$.
In general the integral in equation (\ref{integral}) has no closed form, and it is typically very demanding to compute numerically.  
As an approximation one may consider a second-order Taylor expansion of $\log p(x,h)$ around the maximum 
\begin{equation}
h^*=\op{argmax}_h\log p(x,h).
\label{hmax}
\end{equation}
The resulting integral is easily computed, giving the expression
\begin{equation} \label{Laplace}
p_x(x) \approx \displaystyle (2 \pi)^{T/2}\,|\det \Omega(x)|^{-1/2}\,p(x,h^*)\,,
\end{equation}
where
\begin{equation} \label{Omega}
\Omega(x)=\frac{\partial^2 \log p(x,h)}{\partial h^T  \partial h}\Big|_{h=h^*}
\end{equation}
is the Hessian matrix of the map $h \mapsto \log p(x,h)$, evaluated at $h=h^*$. 
This approximation is known as Laplace's method, and it has been applied, by among others \cite{Martino:2012be}, to compute likelihood functions in basic SV models. 
The reason for its efficiency in basic SV models is the Markov structure of the latent field $h$. The Markov property ensures that the gradient of $h \mapsto \log p(x,h)$ is on the form   
\begin{equation} \label{deriverte}
\frac{\partial \log p(x,h)}{\partial h_i} =b_i + \sum_{j} A_{ij} \,h_j + g_i(x_i,h_i)\,,
\end{equation}
where $A=||A_{ij}||$ is a tridiagonal matrix, $b_i$ are constants and $g_i$ are non-linear functions. By exploiting the sparseness of $A$, one can efficiently calculate $h^*$. In addition, the Hessian matrix $\Omega(x)$ is tridiagonal, making the computation of the expression in equation (\ref{Laplace}) efficient.

We are now in a position to make statistical inference based on the basic SV model. 
We start by looking at filtering of the volatilities. Overlooking model uncertainty and parameter uncertainty, the conditional density $p(h_T|x)$ contains all available information about the latent variable $h_T$ at time $T$. As a point estimate one may consider the Bayes estimator which is defined as the maximum of the posterior distribution $p(h_T|x)$.  This density is approximated using Laplace's method :
\begin{equation} 
\begin{array}{lllll}
\displaystyle p(h_T|x) &\propto& \displaystyle \ p(x,h_T) 
\\
& =&\displaystyle \int_{\mathbb{R}^{T-1}} p(x,h)dh_1\cdots dh_{T-1}\\
&\approx& \displaystyle b\, p\left(x,\tilde{h}_{1},\ldots \tilde{h}_{T-1},h_T\right),
 \,
\end{array} 
\label{filterLaplace}
\end{equation}
where the factor $b$ does not depend on $h_T$, and
\begin{equation}
\left(\tilde{h}_{1},\ldots \tilde{h}_{T-1}\right)=\op{argmax}_{h_1,\ldots, h_{T-1}}\log p(x,h).
\end{equation}
Maximizing (\ref{filterLaplace}) gives the filtered estimate $\hat{h}_T$ of $h_T$. The filtering procedure can be written more compact as in (\ref{hmax}) with $\hat{h}_T=\hat h^*_T$.

For smoothing we consider the posterior distribution $p(h_s|x)$, now with $s<T$. A similar argument as for the derivation of the filtering formula gives the approximated Bayes estimator $\hat{h}_s=h^*_s$ where $h^*_s$ is component $s$ of the vector $h^*$ in equation $(\ref{hmax})$.

We note that if we are already calculating the likelihood using the approach described above, then very little additional effort is required to obtain these estimates, since the maxima in equation (\ref{hmax}) is found as a part of the Laplace approximation. 

To forecast the volatility $N$ steps into the future we follow the same procedure as for smoothing and filtering. We need to find the maximum of the expression 
$$
\log p(x,h,h_{T+N})=\log p(x,h)+\log p(h_{T+N}|h) 
$$
as a function of $(h,h_{T+N})$.  
Iterating equation (\ref{AR}) backwards yields 
$$
h_{T+N}=\psi^N h_T +\sum_{k=0}^{N-1}\psi^k u_{T+N-k}\,,
$$
and hence 
$$
 h_{T+N}|h\dd \N\left(\psi^N h_T,\sigma_u^2 \sum_{k=0}^{N-1}\psi^{2k} \right)\,.
$$
Differentiation of $\log p(h_{T+N}|h)$ gives
\begin{equation}
\frac{\partial  \log p(h_{T+N}|h)}{\partial h_{T+N}} = -\frac{(h_{T+N}-\psi^N h_T)}{\sigma_u^2 \sum_{k=0}^{N-1}\psi^{2k} }\,.
\label{dN}
\end{equation}
To find a maximum we require that the expressions in equation (\ref{dN}) equals zero, and also that 
$\nabla_h \log p(x,h) = 0$. From this, the $N$-step volatility forecast becomes 
\begin{equation} \label{svforecast}
\hat{h}_{T+N}=\psi^N \hat{h}_T,
\end{equation}
where $\hat{h}_T$ is the filtered estimate of $h_T$. The formulas we have derived for smoothing, filtering and forecasting of the volatilities are the same as in \cite{Skaug:2009vna}.

To conclude this section we remark that, since $p(x_{T+N}|x)\propto p(x,x_{T+N})$, the conditional densities $p(x_{T+N}|x)$ can be computed simply by using the Laplace approximation. For $N>1$ one must take into account that the matrices $A$ and $\Omega(x,x_{T+N})$ are modified due to the inclusion of the density $p(h_{T+N}|h)$. 

\noindent
\begin{figure*}[t]
\begin{center}
\includegraphics[width=15.0cm]{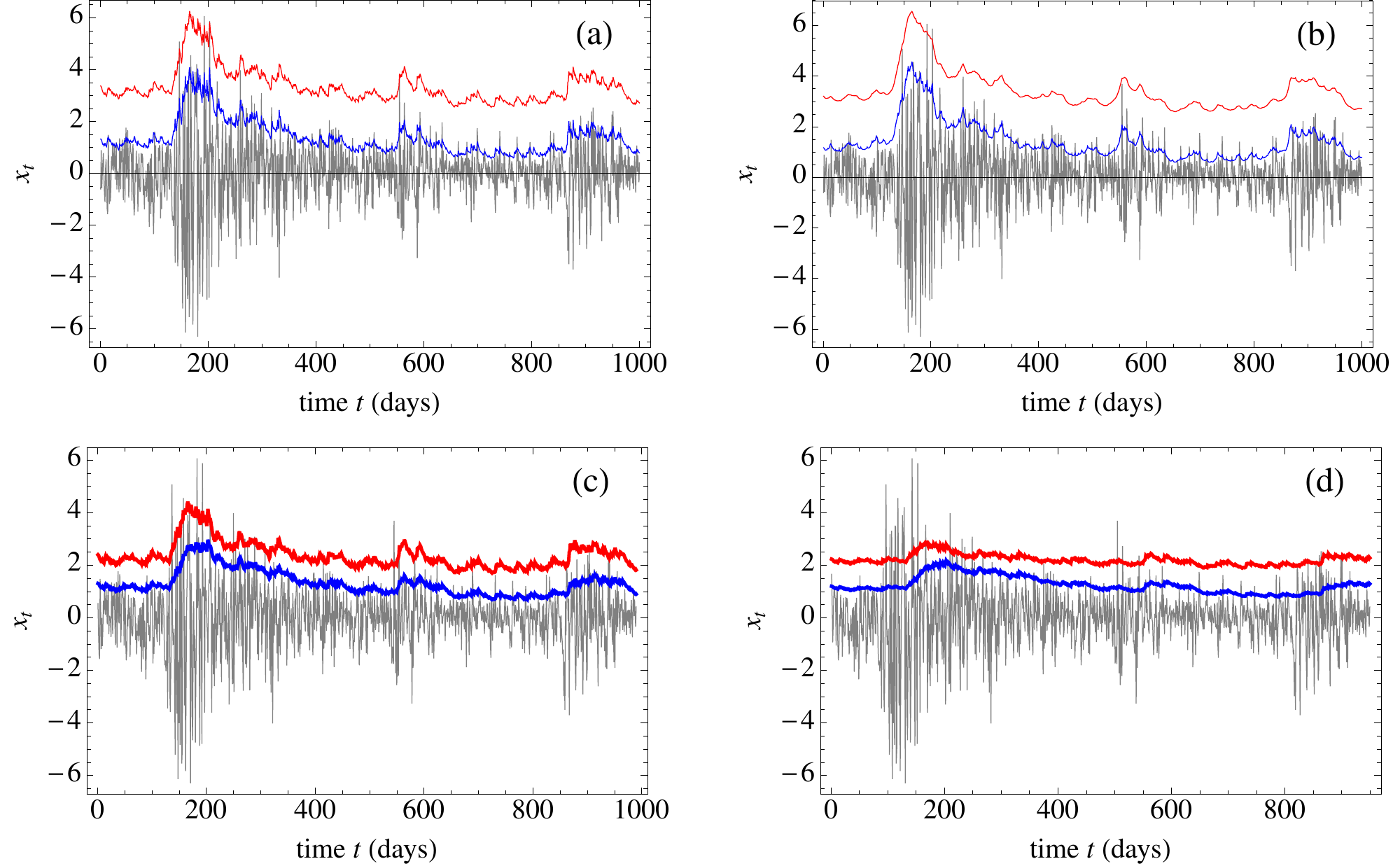}
\caption{(a):  Shows $e^{\hat{h}_t/2}$, where $\hat{h}_t$ are the filtered estimates of $h_t$ from the daily log-returns in the OSEBX in the time period from February 20th 2008 to February 8th 2012. The lower curve is for the MRW model, and the top curve is for the basic SV model. The top curve has been shifted to make it visible. The filtered signals are plotted together with the log-returns $x_t$ of the OSEBX. (b):  Shows the same as in (a), but now $\hat{h}_t$ are the smoothed estimates of $h_t$ given all the observations of $x_t$. (c): Shows $e^{\hat{h}_{t}/2}$, where 
$\hat{h}_{t}$ is the forecast performed using data $\{x_s : s \leq t-N\}$ with $N=10$ days. (d): Shows the same as in (c), but now with $N=50$ days.} \label{fig1}
\end{center}
\end{figure*}
\noindent
\begin{figure*}
\begin{center}
\includegraphics[width=15.0cm]{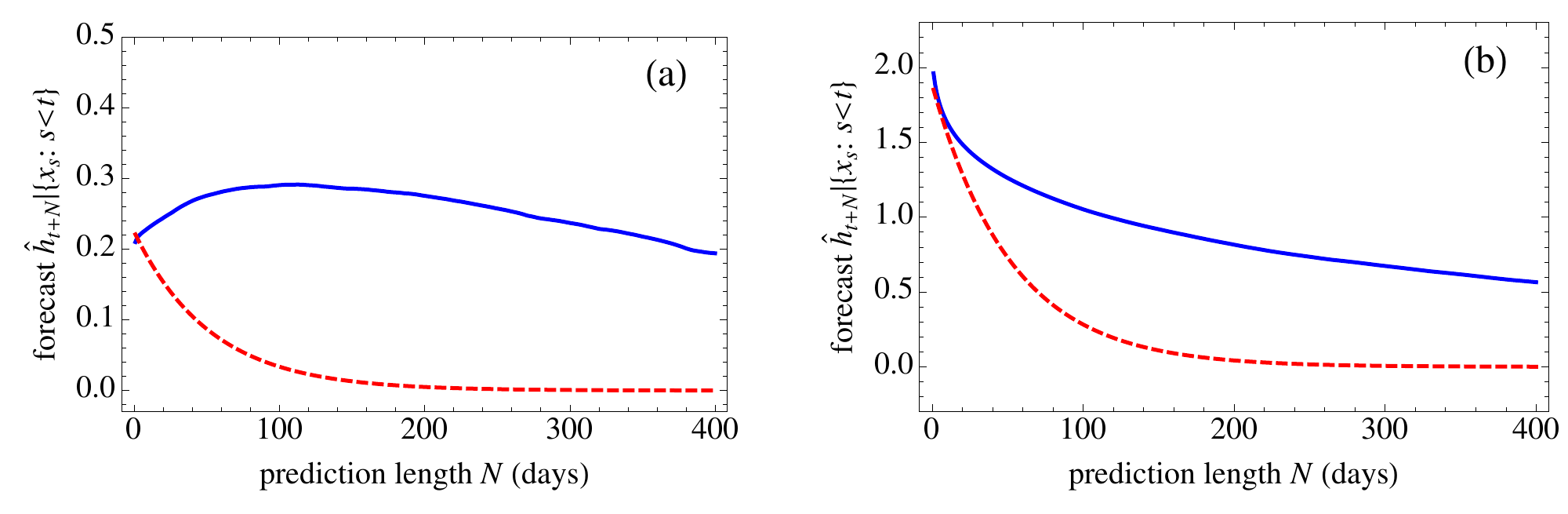}
\caption{(a):  Examples of forecasts $\hat{h}_{t+N}$ for the basic SV model (dotted curve) and the MRW model (solid curve). The forecasts are computed from the OSEBX data with the time $t$ corresponding to the date February 17th 2010. (b): Same as in (a), but now the forecasts are preformed for the date March 5th 2009.} \label{fig2}
\end{center}
\end{figure*}

\section{Generalization to the MRW model} \label{mrwsec}
In this section we extend the results of section \ref{sv} to the discrete-time MRW model. In this case $h_t$ is no longer a Markov process. While $h_t$ still is a centered Gaussian process, its covariance structure is now given by equation (\ref{cov}). 

Let us first review the approximation of the likelihood for the MRW model \citep{Lovsletten:2011vm}. One starts with Laplace's method, given in equation (\ref{Laplace}). The density $p(x|h)$ is the same as for the basic SV model, but the density of $p(h)$ needs to be handled differently. We denote by $\gamma(k)=\op{Cov}(h_0,h_k)$ the auto-covariance function of the process $h_t$, and let  $\Gamma_t$ be the variance-covariance matrices of the vectors $(h_1,\ldots,h_{t})$. That is 
$$
\Gamma_t(i,j)= \gamma(|i-j|) \mbox{~ for ~} i,j=0,1,\ldots,t-1.
$$
As usual when working with Gaussian vectors, it is convenient to introduce regression coefficients $\phi^{(t)}_i$. The vectors $\phi^{(t)}$ are defined via the equations 
\begin{equation} \label{Phi}
\Gamma_{t} \phi^{(t)}  = \gamma_{1:t}\,,
\end{equation}
where $\gamma_{1:t} = (\gamma(1),\ldots, \gamma(t))^T$.
From standard theory of multivariate normal distributions, the conditional distributions of $h_t|\{h_s:1\leq s<t\}$ are normal, 
\begin{equation} \label{betingahmrw}
h_t|\{h_s:1\leq s<t\}\dd\N(m_t,P_t),
\end{equation}
with means $m_t = ( h_{t-1},h_{t-2},\ldots,h_1)\phi^{(t-1)}$ and variances $P_t =\gamma(0)-\gamma_{1:t-1}^T \Gamma_{t-1}^{-1}\gamma_{1:t-1}$. 
Since the density of $h$ can be decomposed into a product of one-dimensional marginals, equation (\ref{betingahmrw}) gives $p(h)$. 

\begin{rk}{\em 
Solving the the equations in (\ref{Phi})  for $t=1,\ldots, T-1$ can be done iteratively using the Durbin-Levinson algorithm, which requires only $\mathcal{O}(T^2)$ floating point operations. We refer to 
\citep{McLeod:2007wp} for details.}
\end{rk}

A second difference between the basic SV model and the MRW model is the structure of the matrices $\Omega$  and $A$, which are defined by equations (\ref{Omega}) and (\ref{deriverte}). For the MRW model these are no longer sparse. This makes the computation of the expression in equation (\ref{Laplace}) extremely demanding. The solution is to truncate the dependency in the process $h_t$ after a finite number of lags. This gives the approximation:
\begin{equation}
p(h_t|\{h_s:1\leq s<t\}) \approx p(h_t|\{h_s:t-\tau \leq s<t\})\,,
\end{equation}
where $\tau \in \mathbb{N}$ is a truncation parameter. 
We note that for $t>\tau$, the regression coefficients and variances of $h_t|\{h_s:t-\tau \leq s<t\}$ are $\phi^{(\tau)}$ and $P_{\tau+1}$ respectively. After truncation, the matrices $A$  and $\Omega$ become band-diagonal with bandwidths equal $\tau$.

\begin{rk} {\em The likelihood approximation for the MRW model is implemented in the R computer language. In our implementation we have used analytical expressions for the first and second order derivatives to construct the matrices $\Omega$ and $A$. The maxima $h^*$ are found by numerically calculating the roots of the expressions in equation (\ref{deriverte}) using the algorithm "DF-SANE" \citep{LaCruz:2006vo}. This algorithm is implemented in the R package "BB" \citep{Varadhan:2009vb}. To find the determinant of the $\Omega$ we use the package "Matrix" which efficiently stores and manipulates sparse matrices. }
\end{rk}

With the likelihood approximation at hand, we can extend the formulas for smoothing, filtering and forecasting to the MRW model. As for the basic SV model, we  maximize the posterior distribution according to equation (\ref{hmax}), and the formulas for smoothing and filtering are exactly as for the basic SV model. 

To forecast the volatilities $N$ steps ahead we need the conditional density of $h_{T+N}|h$. Since this variable is normal, the distribution is uniquely given by the mean $m_{T+N|T}$ and variance $P_{T+N|T}$, i.e.
\begin{equation}
h_{T+N}|h\dd \N(m_{T+N|T},P_{T+N|T}).
\end{equation}
The mean is a linear combination the conditioning variables, i.e. 
$$
m_{T+N|T}= ( h_{T},h_{T-1},\ldots,h_1)\phi^{(T,N)}\,,
$$ 
where the coefficients $\phi^{(T,N)}$ are solutions to the equations
\begin{equation}
\Gamma_T \phi^{(T,N)}= \gamma_{N:T+N-1}\,, 
\end{equation}
with
$$
\gamma_{N:T+N-1} =  (\gamma(N), \gamma(N+1),\ldots, \gamma(N+T-1))^T\,.
$$
The variance is given by 
$$
P_{T+N|T}=\gamma(0)-\gamma_{N:T+N-1}^T \Gamma_T\gamma_{N:T+N-1}\,.
$$ 
We note that in the special case $N=1$ we have $\phi^{(T,N)} =  \phi^{(T)}$, and we can again use the Durbin-Levinson algorithm. 
In the case $N > 1$, the explicit inverse of $\Gamma_T$ is needed, and one may use the algorithm of \cite{Trench:1964vj}, which utilizes that the matrices are Toeplitz. Using the same procedure as in section \ref{sv}, we get the forecasting formula 
\begin{equation} \label{mainresult}
\hat{h}_{T+N}=\left( \hat{h}_{T},\hat{h}_{T-1},\ldots,\hat{h}_1\right)\phi^{(T,N)}\,,
\end{equation}
where $\left(\hat{h}_{1},\hat{h}_{2},\ldots,\hat{h}_T\right)$ are the smoothed estimates of (${h}_{1},{h}_{2},\ldots,{h}_T$).

Using the Laplace approximation for the MRW model, the $N$ step conditional densities $p(x_{T+N}|x)$ are computed as in section \ref{sv}.

\noindent
\begin{figure*}[t]
\begin{center}
\includegraphics[width=15.0cm]{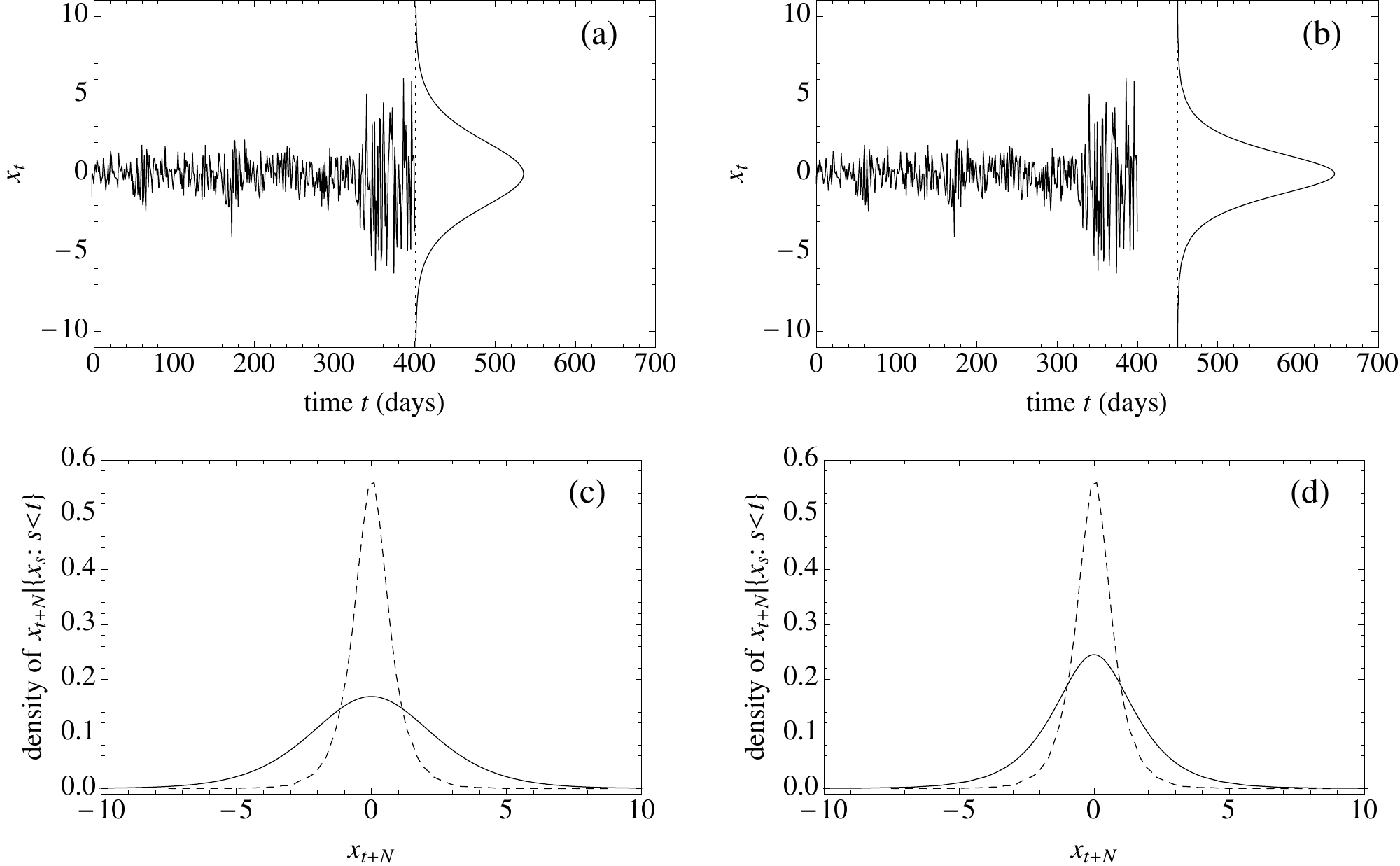}
\caption{(a): An illustration of the use of conditional densities in forecasting. The figure shows the log-returns of the OSEBX up to December 12th 2008 and the conditional density of the log-return for the next day. (b): Shows the same as (a), but now the conditional density is a $N$-step forecast with $N=50$ days. (c): Shows the same conditional density (solid curve) as is illustrated in (a). This density is compared with the unconditional density for the log-returns (dashed line). (d): Same as in (c), but now for the $N$-step forecast with $N=50$ days} \label{fig3}
\end{center}
\end{figure*}
\section{Examples} \label{example}
As an example we have applied the inference methods presented in sections \ref{sv} and \ref{mrwsec} to a time series consisting of daily log-returns of the Oslo Stock Exchange Benchmark Index (OSEBX). The data used are closing prices for the time period May 25th 2001 to February 8th 2012, and the whole time series is used to obtain ML estimates for the basic SV model and the MRW model. The interesting estimates are $\hat{\psi} = 0.98$ for the basic SV model and $\hat{\lambda}  = 0.33$ for the MRW model.     

In figure \ref{fig1}(a) we have plotted the filtered estimates of $h_t$ together with the log-returns for the time period from February 20th 2008 to February 8th 2012. The filtering for the basic SV model and the MRW model are similar, but not identical. The same is seen in figure \ref{fig1}(b), which shows the smoothed estimates. In figures \ref{fig1}(c) and \ref{fig1}(d) we have plotted the $N$-step forecast for $N=10$ days and $N=50$ days respectively. In the 50-day forecast there are some clear visible differences between the two models. These differences become even clearer in figure \ref{fig2}. In this figure we show two examples, where we (for a fixed time $t$) make future predictions $\hat{h}_{t+N}$, and plot these as a functions of $N$. It follows from equation (\ref{svforecast}) that these curves must be monotonic and exponentially decaying for the basic SV model. This is not the case for the MRW model, and we observe that the forecasts based on the this model have much richer behavior. We note that similar observations have been made for the MSM model \citep{Calvet:2001cw}.    

As explained in sections \ref{sv} and \ref{mrwsec}, it is possible to use the Laplace approximation to compute the full conditional densities for future returns. This gives forecasts containing more information than the estimates presented in figures \ref{fig1} and \ref{fig2}. In figure \ref{fig3} we show two examples where such densities have been computed. In these examples, the volatility is high, and the conditional densities are wider than the unconditioned density. In other situations, where the volatility is low, the conditional densities will be narrower than the unconditioned density.  

\begin{rk}{\em 
The computer code that is used for these examples is available online at \url{complexityandplasmas.net}}. 
\end{rk}

\section{Conclusion} \label{conclusion}
The main results of this paper are methods for smoothing, filtering and forecasting using the MRW model. In addition, we have presented methods for computing conditional densities of future returns. These results improve on existing forecasting techniques for multifractal models, and we therefore consider this work to be an important contribution to the field. 

The methods presented in this work open the way for several future studies of multifractal modeling in finance. Among the new possibilities that we consider most interesting, are model comparisons based on estimated future distributions.   

\section*{Acknowledgement}
\noindent This project is supported by {\em Sparebank 1 Nord-Norge}.

\appendix

\section*{References}
\bibliographystyle{elsarticle-harv}
\bibliography{ref.bib}

\end{document}